\documentclass{llncs}
\usepackage{cite}
\usepackage{graphicx}
\usepackage{url}

\begin{document}

\title{Decomposition of quantitative Gaifman graphs \\ 
as a data analysis tool\thanks{%
This research was supported by grant 
TIN2017-89244-R from Ministerio de 
Economia, Industria y Competitividad, and by
Conacyt (M\'exico); and
we acknowledge recognition 2017SGR-856
(MACDA) from AGAUR (Generalitat de Catalunya).%
}}

\author{Jos\'e Luis Balc\'azar\inst{1} \and Marie Ely Piceno\inst{2} \and Laura Rodr\'iguez-Navas\inst{3}} 
\institute{Universitat Polit\`ecnica de Catalunya}
%\email{jose.luis.balcazar@upc.edu  mpiceno@cs.upc.edu}

\maketitle

\begin{abstract}
We argue the usefulness of Gaifman graphs 
of first-order relational structures as an exploratory 
data analysis tool. We illustrate our approach
with cases where the modular decompositions of 
these graphs reveal interesting facts about the data.
Then, we introduce generalized notions of Gaifman
graphs, enhanced with quantitative information, 
to which we can apply more general, existing 
decomposition notions via 2-structures; thus
enlarging the analytical capabilities of the scheme.
The very essence of Gaifman graphs makes this approach
immediately appropriate for the multirelational
data framework. 
\end{abstract}

\section{Introduction}

% \looseness=-1
First-order (finite) relational structures 
(see e.g.~\cite{DBLP:books/sp/Libkin04})
are the conceptual
essence of the relational database model. 
Gaifman graphs are a well-known, quite natural theoretical
construction that can be applied to any
relational structure~\cite{DBLP:books/sp/Libkin04}. 
They have provided very interesting progress 
in the theory of these logical models. 

Given a first-order relational structure, or
relational database, with relations (or tables)
$R_i$, where the values in the tuples come from 
a fixed universe $U$, the corresponding Gaifman 
graph has the elements of $U$ as vertices; and
there is an edge $(x,y)$, for $x\neq y$, exactly 
when $x$ and $y$ appear together in some 
tuple~$t\in R_i$ for some table~$R_i$.
That is, Gaifman graphs record co-occurrence (or lack
thereof) among every pair of universe items.

Hence, a clique in a Gaifman graph would group items
that, pairwise, appear together somewhere in 
the relational structure: co-occurrence patterns; 
a clique in its complement would reveal an 
incompatibility pattern. Of course, finding 
maximal cliques is NP-complete; but there are 
less demanding ways to study graphs that identify 
efficiently both sorts of patterns in a recursive 
decomposition: namely, the modular decomposition and 
its generalization, the decomposition of 2-structures.

This paper proposes to employ these decompositions as
avenues for exploratory data analysis on 
relational data (whether single- or multi-relational):
by applying them on the Gaifman graph of a
dataset, we can obtain valuable information
that would not be readily observable directly
on the data.

Modular decompositions suffice to treat 
stardard Gaifman graphs. However, we extend 
the capabilities of our approach 
by generalizing, in very natural ways, the 
notion of Gaifman graph so as to handle 
quantitative information (a must in many
data analysis applications). Hence, we
develop our work using the more general
decomposition of 2-structures~\cite{DBLP:books/daglib/0025562}:
again a notion
that has been very fruitfully developed in 
their theoretical form, and in a number of
applications (such as \cite{DLarraz2010}), 
but not yet imported, to our knowledge, 
into data analysis frameworks.

\section{Decomposing standard Gaifman graphs}
\label{s:decomp}

As already mentioned, 
the basic notion of Gaifman graph is pretty simple:
on all items that appear along all the tuples of
a single- or multi-relational dataset, edges join
pairs of items that appear together in some tuple.

\begin{figure}
\vglue-8mm
\centering
\includegraphics[width=0.26\textwidth]{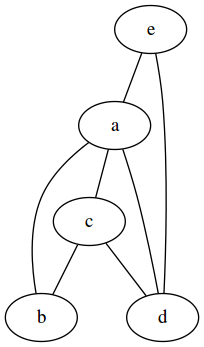}
\hskip13mm
\includegraphics[width=0.25\textwidth]{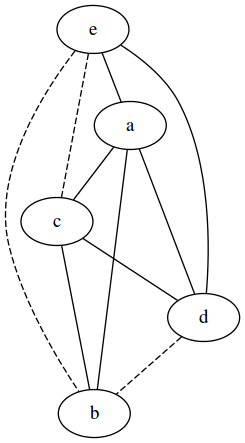}
\hskip13mm
\includegraphics[height=0.33\textheight]{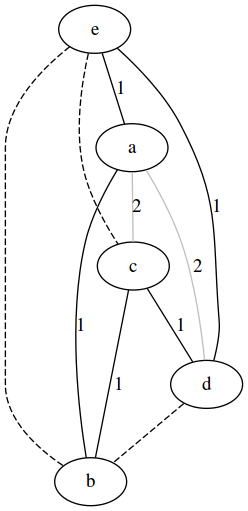}
\caption{A Gaifman graph, its natural completion, and a labeled variant.}
\label{fg:easygg}
\end{figure}

\begin{example}
\label{ex:easygg}
As a running example, let us consider a very small, 
single-relation database on the
universe $\{a, b, c, d, e\}$,
with three attributes and three tuples:

% exampleasy:

\qquad $t_1$: $a$ $b$ $c$

\qquad $t_2$: $a$ $d$ $e$

\qquad $t_3$: $a$ $c$ $d$

Then, the Gaifman graph is as shown in Figure~\ref{fg:easygg} (left).
\end{example}

%\begin{figure}
%\centering
%\includegraphics[height=0.25\textheight]{img/exampleasygg.png}
%\caption{Gaifman graph for Example~\ref{ex:easygg}.}
%\label{fg:easygg}
%\end{figure}

\subsection{2-structures and their decompositions}

\looseness=-1
The very classical notion called 
``modular decomposition''~\cite{Gallai1967} 
suffices to implement our approach on plain Gaifman graphs;
this notion has been rediscovered many times and described
under several different names\footnote{See 
\url{https://en.wikipedia.org/wiki/Modular_decomposition} 
for some of the alternative names that the concept has received.}.
However, it is insufficient to handle adequately the generalizations that we 
will
propose below.
Therefore, 
we develop our approach
directly on top of the more general notion of 2-structures 
and their clans~\cite{DBLP:books/daglib/0025562}.

First, we describe some ``cosmetics'' on our Gaifman graphs:
they will be seen 
as a complete graph with 
two sorts of (nonreflexive) edges.
One sort corresponds to edges present in
the graph (solid lines in our
diagrams); the other corresponds to absent 
(nonreflexive) edges
(broken lines).
We call this graph the ``natural completion''
of the original graph. In our example,
this process is illustrated in Figure~\ref{fg:easygg} (center). % {fg:easycompgg}

Additionally, we can label each
edge with its multiplicity, that~is, 
the number of tuples that contain
the pair of items linked by the edge. 
The previous example then becomes as in 
Figure~\ref{fg:easygg} (right): pairs appear either
zero times together (dashed edges), once
(black lines, labeled 1) or, in two cases,
twice (gray lines, labeled 2). 

Now, in general terms,
a 2-structure is simply the complete graph on some universe $U$,
plus an equivalence relation $E$ among the edges.
Figure~\ref{fg:easygg} (right) serves as 
an example, where there are three equivalence
classes of edges: the broken edges, the black
edges, and the gray edges;
% (and ignoring the numeric labels); 
of course, Figure~\ref{fg:easygg} (center) % {fg:easycompgg}
is also an example, with just two equivalence
classes of edges.
We will restrict ourselves to undirected
edges, and will employ the common, very graphical and
intuitive representation of coloring in the
same way edges belonging to the same equivalence
class. 

\looseness=-1
Observe that the type of the equivalence relation $E$ is 
$E\subseteq ((U\times U) \times (U\times U))$
because $E$ tells us whether two
arbitrary edges $(x,y)$ and $(u,v)$ are equivalent.

For a 2-structure given by the set of vertices $U$
and the equivalence relation~$E$ among the edges
of the complete graph on $U$, we say that a subset
$C\subseteq U$ is a clan, informally, if all the members of $C$
are indistinguishable among them by non-members. That is:
whenever some $x\notin C$ ``can distinguish'' between $y\in C$
and $z\in C$, in the sense that the edge $(x,y)$
is not equivalent to the edge $(x,z)$, then $C$
is \emph{not} a clan. Formally (see~\cite{DBLP:books/daglib/0025562}):

\begin{definition} 
Given $U$ and an equivalence relation
$E\subseteq ((U\times U) \times (U\times U))$
on the edges of the complete graph on $U$,
$C\subseteq U$ is a clan when
$$
\forall x\notin C \,\forall y \in C \,\forall z \in C \,((x,y),(x,z))\in E.
$$
\end{definition} 

Note that different vertices outside the clan 
might see the clan differently: for $x\notin C$
and $x'\notin C$, and $y\in C$, the edges $(x,y)$
and $(x',y)$ may well be nonequivalent. We only
require that each fixed $x$ does not distinguish
between the clan members.

Basic examples of clans are the so-called trivial clans:
all the singletons $\{x\}$ for $x\in U$, as well
as $U$ itself, are vacuously clans. There may
be other clans. For instance, consider the natural
completion of the Gaifman graph obtained from 
Example~\ref{ex:easygg}, 
depicted in Figure~\ref{fg:easygg} (center). % {fg:easycompgg}
Edges are split into two equivalence classes
(existing or nonexisting edges in the original
Gaifman graph). Then, one can see that there 
would be exactly one nontrivial clan, formed 
by $\{b, c, d, e\}$: all vertices not in the clan 
(that is, vertex~$a$, the single one not in the clan) 
are connected to each vertex inside the clan through 
edges of the same color, namely solid black. Any other
candidate turns out not to be a clan. For instance,
any set including $a$ and $b$ but excluding $e$
is not a clan, as $e$ ``distinguishes'' between
$a$ and $b$; then, any set including $b$ and $e$
must include $c$ and $d$, which can distinguish
between them. All in all, any clan including $a$
and $b$ ends up including all the vertices, that
is, becoming a trivial clan. Analogous reasoning
applies if we start by pairing $a$ with other vertices.

On the other hand, it is not difficult to see that the
labeled, colored version of the Gaifman graph
of Example~\ref{ex:easygg}, as depicted in 
Figure~\ref{fg:easygg} (right), does not have nontrivial 
clans. Equivalence is given by the same
multiplicity label (that is, edges drawn 
in the same ``color''): the extra distinction 
between gray and
black edges allows for external vertices to 
distinguish between some vertices inside every 
candidate proper subset. Further 
examples come later as clans are the key tool
for our proposal of a data analysis method.

\subsection{Prime clans and tree decompositions}

It is known~\cite{DBLP:books/daglib/0025562} that 
certain clans, called prime clans, allow us to 
decompose a 2-structure into a tree-like form. 

\begin{definition}
For a fixed universe $U$, we say that two
subsets of $U$ overlap if neither is a subset
of the other, but they are not disjoint.
That is, for $S\subseteq U$ and $T\subseteq U$,
they overlap if the three sets $S\cap T$, 
$S\setminus T$, and $T\setminus S$ are 
all three nonempty.
Then, prime clans are those clans that do not
overlap any other clan. 
\end{definition}

Of course, trivial clans are also immediately
prime clans. Thus, by definition,
any two sets in the family of prime clans 
are either disjoint, or a subset of one another: they 
provide us with a so-called
``decomposable set family''~\cite{MCCONNELL1999189}
 that can be pictured 
in a tree form, by displaying every prime clan (except $U$
itself) as a child of the smallest prime clan
that properly contains it. 

There are studies that report how these
decompositions look like. Specifically, 
% we easily see that 
at each node of the tree we have again a 2-structure, 
whose vertices correspond to the clans that fall as children of the
node. In the case of our constructions out of Gaifman
graphs, it is known that all the 2-structures
that appear as nodes of such a tree decomposition are
of one of two well-defined sorts: either ``complete''
(all edges are equivalent) or ``primitive'' (only having
trivial clans). This is due to our graphs being undirected,
because 2-structures on directed graphs may exhibit a third 
basic component in their tree decomposition (``linear''
2-structures). Further information
on this topic appears in~\cite{DBLP:books/daglib/0025562}. 
This reference contains, as well, often far-from-trivial
proofs of theorems that ensure that things are as
we have described.

\begin{example}
\label{ex:easyggdecomp}
Continuing Example~\ref{ex:easygg}, 
the tree decomposition of the 2-structure in 
Figure~\ref{fg:easygg} (center) is displayed in  % {fg:easycompgg}
Figure~\ref{fg:easydec} (left). Boxes correspond to
clans: here, the topmost box
corresponds to the trivial clan containing all
the vertices and, inside it, each dot corresponds to a 
prime subclan. All along the whole decomposition, trivial clans 
are indicated by a link
to the vertex they consist of, represented with 
an elliptic node; nontrivial ones
are linked instead to a new box describing the internal
structure of the clan, in terms of the prime clans 
it has as proper subsets. Then, as a set, each clan is
formed by all the elements in the leaves of
the subtree rooted at it. 
\end{example}

A % first 
``brute-force'', exhaustive search attempt 
was employed in \cite{LRodriguez2017} to identify
all prime clans.
A couple of published algorithms~\cite{MCCONNELL1999189, DBLP:journals/algorithmica/McConnell95} 
can be adapted 
for implementing a system computing this sort of
tree decompositions. However, as we envision an
analysis support system able to add Gaifman nodes
in an incremental manner, we have implemented a somewhat
different, incremental algorithm. Due to the space limit,
the details of our algorithmic contributions will be presented
in a follow-up paper (or in an expanded version of this one), 
together with some comparisons against other algorithms.

\subsection{Limits to the visualization of complex clans}
\label{ss:limits}

Our experimentation shows that, unsurprisingly,
the visualization of large Gaifman graphs is unadvisable.
Actually, sometimes the clans lead to large primitive 
2-structures, whose mathematical study gets pretty 
complicated~\cite{DBLP:journals/tcs/EhrenfeuchtR90b}.
We set up some relatively arbitrary limits, trying to
get understandable diagrams. Let us consider a more
realistic example to explain them.

\begin{figure*}
\vglue-3mm
\centering
\includegraphics[height=0.28\textheight]{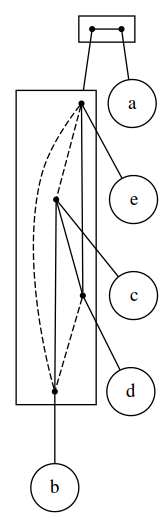}
\hskip 2cm
\includegraphics[height=0.28\textheight]{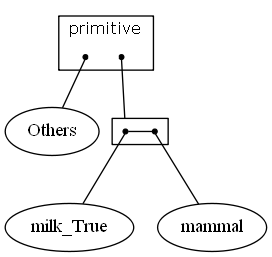}
\caption{Decompositions of the Gaifman graphs for Example~\ref{ex:easygg} and for the Zoo dataset.}
\label{fg:easydec}
\end{figure*}

% \begin{example}
In Figure~\ref{fg:easydec} (right) we display 
(a fragment of) the
decomposition of the Gaifman graph of the well-known
Zoo dataset from the UC Irvine repository
\cite{UCI}; it contains 17 attributes of 100
animal species. We have preprocessed it slightly so
that the semantics of each item is clearly identifiable
(e.~g.~predator\_False or toothed\_True). We will return
to this dataset below in Section~\ref{ss:quantgg}.

For the time being, we just discuss the decomposition
of its standard Gaifman graph. The topmost node 
of this decomposition is, as always, the trivial clan
with the whole universe; in this case, it turns out 
to decompose as a set of many trivial clans, set up 
in the form of a primitive 2-structure that we
choose not to draw complete; however,
one nontrivial clan also appears: ``mammal'' and
`milk\_True'' are indistinguishable from the
perspective of all the other elements in the dataset.
That is, for every other piece of information, either it
goes together with each in some tuples (one such item 
could be ``hair\_True''), or it does not go together with 
any of them ever (for instance: ``feathers\_True'').

In our diagrams, as we do here, 
clans containing more than a handful of nontrivial clans are not 
drawn in detail: just the clan type label (``primitive'' 
or ``complete'') is shown. Besides, if there are few
nontrivial clans, but many trivial ones,
then the trivial clans are grouped in a single 
node labeled Others, sort of a merge of them all. 
The reader must % is to 
keep
in mind that this particular node actually
represents together a number of unstructured items.

%  as with the isolated vertices;
% and we draw the nodes that represent the other clans
% together with the clan type label. Clans conformed by 
% less than six smaller clans, whether trivial or nontrivial, 
% are fully displayed. 
This approach of limiting the size
of the substructures that become fully spelled-out was
taken also in \cite{DLarraz2010}, where also a ``zooming''
capability was introduced (we may consider adding one
such option to our system in the future).

\subsection{Isolated vertex elision}
\label{ss:elision}

As we move on, later, into quantitative
generalizations of Gaifman graphs,
one case turns out to be common in our experiments. 
Whereas Gaifman graphs do not have isolated
vertices (except in limit, artificial cases such as
relations with a single attribute), in
our generalizations this is no longer true: 
many datasets will lead to 2-structures exhibiting many  
vertices that are endpoints only of broken
edges; that is, they are isolated vertices 
in the corresponding (generalized) Gaifman graph. 
The set of those 
isolated vertices forms a sometimes quite 
large clan that clutters the
diagram but contributes nothing to the analysis
beyond ``all these vertices are actually isolated''.
We use again the label ``Others''
to represent these items, all alike from the 
decomposition perspective, as a single
vertex, as indeed this is a particular case
of the usage of the ``Others'' label as per
the previous Section~\ref{ss:limits}.

% \subsection{Algorithmics and implementation}

\section{Interpreting a decomposition of a Gaifman graph}

We move on to explain another example of our analysis strategy.
We present and discuss the outcome of a tree decomposition
of the Gaifman graph of a simple, famous, and relatively small 
dataset often used for teaching introductory data analysis
courses. It comes from data of each of the passengers of the
Titanic. Among several existing variants of this dataset, some
of them pretty complete, we choose a reduced variant on which
we illustrate the interpretation of our decompositions.
This variant we use keeps four attributes, one of them
(age) discretized. To describe the details of this dataset,
we quote:

\smallskip
``The titanic dataset gives the values of four categorical 
attributes for each of the 2201 people on board the Titanic 
when it struck an iceberg and sank. The attributes are social 
class (first class, second class, third class, crewmember), 
age (adult or child), sex, and whether or not the person survived.''

\smallskip
\rightline{(\url{http://www.cs.toronto.edu/~delve/data/titanic/desc.html})}

\smallskip
(As indicated in that website, this variant of the
data was originally compiled by Dawson \cite{Dawson1995}
and converted for use in the DELVE data analysis
environment by Radford Neal.)

\begin{figure*}[!t]
\centering
\includegraphics[height=0.30\textheight]{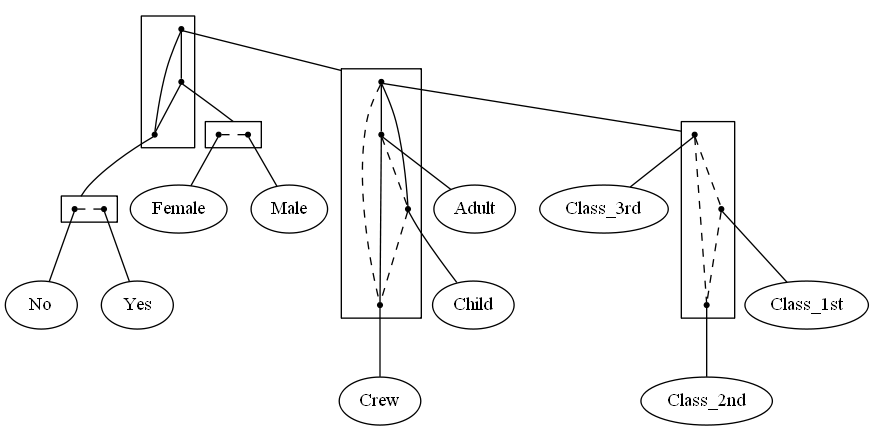}
\caption{Decomposing the standard Gaifman graph of the Titanic dataset.}
\label{fg:titanicplain}
\end{figure*}

The decomposition via its standard Gaifman graph is 
depicted in Figure~\ref{fg:titanicplain}. 
Recall that broken edges represent pairs that never
appear together in any tuple, whereas solid edges are
edges of the original Gaifman graph and thus join
universe elements that appear together in some tuple.

The clans for sex and survival are clear and intuitive: 
as they are different possible values for the same attribute, 
they never appear together. On the other hand, every
possibility for these attributes does appear somewhere,
as does every possible pairing with all other items in
the universe, so that the top node is a complete 2-structure 
consisting on all solid edges.

Likewise, one might expect a clan with the four 
alternative values of traveling class, namely, 
1st, 2nd, 3rd or Crew. However, that clan only has
actual passenger classes. The value Crew 
migrates to the parent ``ages'' clan, 
where we find some interesting fact: 
a small primitive 2-structure arising from
the interaction of the ages values and 
the Crew value, where of course being an adult
is incompatible with being a child, and both
are compatible to all the traveling classes
(the top node in the middle clan); however,
being in the crew is only compatible with
being an adult.
This calls our attention to the fact that 
the crew included, of course, no children, a fact that 
we might overlook in a non-systematic analysis.
That is: even if the traveling classes 
and the ``Crew'' label are
employed as values in the same column, 
the data tells us,
through our decomposition procedure,
that they have different semantics!

\section{Generalizations of Gaifman graphs}

We move on to discuss tree decompositions
based on generalized Gaifman graphs.
The aim is to keep track of quantitative information
that the standard Gaifman graph lacks.
In our context, many ideas present themselves to
complement Gaifman graphs and clan decompositions
with quantitative considerations. For the time
being, we contemplate just some very simple cases: 
we let the number of occurrences of each pair 
play a role.

\subsection{Thresholded Gaifman graphs}

Our first variant is as follows.

\begin{definition} 
For a threshold $k$ (a nonzero natural number)
a \emph{thresholded Gaifman graph} is a 
completion of a Gaifman graph in which each 
labeled edge is classified according to its 
number of occurrences, as follows.
We still have two equivalence classes of edges.
If the number in the label is above the threshold $k$,
the edge goes into one equivalence class
(represented in our diagrams by a solid line);
whereas if the number of occurrences of the edge 
is less than or equal to the threshold, then
the edge belongs to the other equivalence class
(and a broken line is used to represent it). 
\end{definition} 

\begin{figure}
\vglue-3mm
\centering
\includegraphics[width=0.21\textwidth]{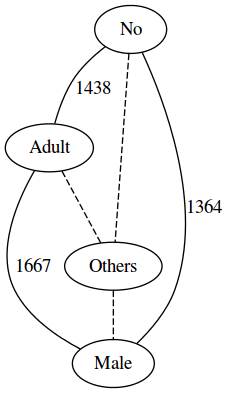}
\hskip 2cm
\includegraphics[width=0.21\textwidth]{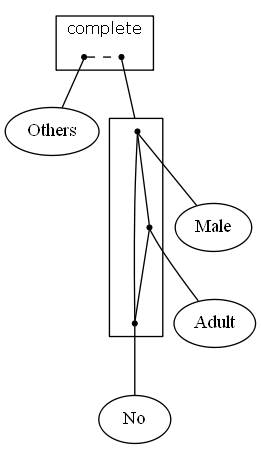}
\caption{Titanic dataset: thresholded Gaifman graph, at 1000, and its decomposition.}
\label{fg:titanicthreshold}
\end{figure}

Figure~\ref{fg:titanicthreshold} provides an alternative
analysis of the Titanic dataset described before. 
There, we decompose a thresholded Gaifman graph, 
aiming at uncovering very common co-occurrences, 
that is, high multiplicities. We set
the threshold rather arbitrarily at 
the quite high value of 1000 (out of 2201 tuples).
We see at work the effect of isolated vertex elision, as many
attribute values to not reach multiplicity 1000 with
any other value: the elision process, as described in
Section~\ref{ss:elision}, replaces all of them by a single
node, playing the same role all of them play, that is,
broken lines among themselves and to all the surviving
values. The new decomposition is interesting in that
it very clearly reflects the Birkenhead Drill:
``Women and children first''.

\subsection{Quantitative Gaifman graphs}
\label{ss:quantgg}

The \emph{linear colored Gaifman graph} is a 
(completion of a) Gaifman graph in which the equivalence
classes of the edges are directly defined by the label,
that is, the number of occurrences. All pairs occurring
once would lead to one class, those occurring twice to
another, and so on; up to some limit, beyond which we do not
keep the distinction. 
Figure~\ref{fg:easygg} (right) corresponds to this case.

\begin{figure}
\centering
\includegraphics[height=0.30\textheight]{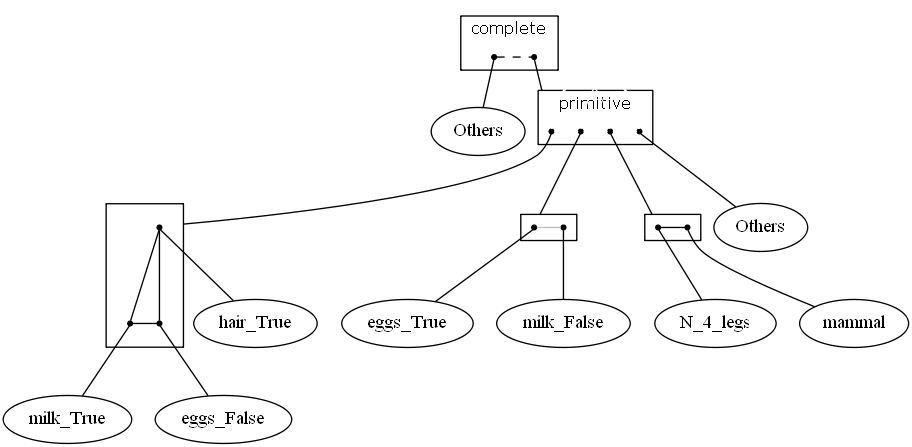}
\caption{Decomposing the linear Gaifman graph of Zoo with 25 as interval width.}
\label{fg:zoolinearcolor}
\end{figure}

%\looseness=-1
A natural variation is to have each color 
stand for some interval of values, with
linearly growing limits; 
the case just described would correspond to intervals
of width~1. 
%As an example, 
Figure~\ref{fg:zoolinearcolor} shows one such case:
we apply intervals of width 25 over the Zoo dataset.
Broken lines mark less than 25 occurrences, solid lines
less than 50, and the gray line appearing inside one of 
the clans goes beyond that limit because it gathers 
all birds and all fish and all insects into the oviparous clan.

%\begin{figure*}[!t]

This notion can be combined
naturally with the previous one: 
instead of broken lines being simply the first interval,
we can apply a different value as threshold
and leave as broken lines all occurrence multiplicities
below it, and then use the colors for the values at the
threshold or above it, at linearly growing intervals
of fixed width. Likewise, an upper threshold can be imposed.
For instance, on the Titanic data, we used colors by
width 1 intervals up to an upper threshold of 10: this 
approach is able to point out for us,
with no particular user guidance, the fact that
the number of children among the first-class 
travelers was surprisingly small: as it happened to
Crew, the first-class node migrates from the
traveling-class clan to the age clan.

\looseness=-1
We expect usefulness also from the
\emph{exponential colored Gaifman graph}: 
while similar in spirit to the 
intervals in linear graphs, 
here
the interval width grows exponentially: each equivalence
class (or color) represents an exponentially 
growing interval of occurrence multiplicities. 
On one hand, this frees the user from having to
bet on a specific interval width. On the other,
there are cases where the Gaifman graph 
multiplicities turn out to be approximately
Zipfian, and the exponential coloring is likely
to be adequate.
Again, as with the linear case, we can also impose a
user-defined threshold below which, or over which, 
the occurrences are not considered different;
then, one runs the exponential count between them.

% \subsection{The multirelational perspective}

Even though the black-and-white printed version
of this paper will not show it, we chose to provide
an example of application of the exponential graph to the 
(``people'' table within the)
UW-CSE dataset from the Relational Dataset
Repository %\footnote{
(\url{http://relational.fit.cvut.cz}) %}
at threshold 3. 
The items have been renamed for better understanding; also,
we have manually edited out a small part of the diagram
to fit the page size and to focus on the three different
colors in the pairs of equivalent items: these colors
tell us that the amount of Students (and thus NotProfessors)
is largish (specifically 216), the quantity of year zero
cases clearly smaller (namely 138) and the amount of
Professors even smaller (62 in total).

\begin{figure*}%[!t]
\centering
\includegraphics[height=0.2\textheight]{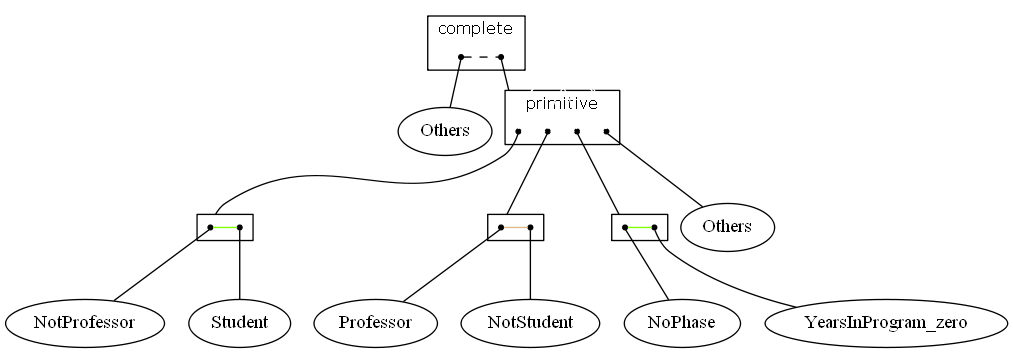}
\caption{UW-CSE: % multirelational dataset: 
part of the decomposition via the 
% thresholded 
exponential Gaifman graph.}
\label{fg:uwexponential}
\end{figure*}

\section{Discussion and subsequent work}

\looseness=-1
We have described a data analysis approach based on
the prime tree decomposition of variations of the
Gaifman graph of a dataset. 
We have illustrated
the process with some relatively successful cases.
Technologically,
we have resorted to a relatively simple implementation
in Python, \url{https://github.com/MelyPic/PrimeTreeDecomposition}, 
% compatible with both version 2 and version 3, and CHECK OUT THIS!!!!
relying on the standard graph module NetworkX and
on the graphical capabilities of the {\tt pydot} interface 
to the Dot engine of 
Graphviz~\cite{Graphviz}.
% Regarding further work, there is a long route ahead.
% First, 
We have not compared 
the available algorithms: there
is no room left 
for that study in this submission,
and it will be the subject of forthcoming write-ups.

% we must point out that finding out
% how to configure it to produce acceptable output was
% a nontrivial task. 

% Some further observations are in order, 
% so as to provide additional perspective.

Many possibilities of further development remain.
First and foremost, we must discuss a clear limitation.
Like in so many other exploratory data analysis frameworks,
for a given dataset we may not be lucky: it may happen that
a given selection of Gaifman graph, once decomposed, has
no nontrivial clans, or decomposes into just a few quite
large primitive substructures that provide little or no
insight about the data. For one, the linear Gaifman graph 
of the well-known toy dataset Weather (discussed 
e.g.~in~\cite{DBLP:books/lib/WittenFH11}) has only trivial 
clans and, if fully displayed, leads to just a large box 
of colored spaghetti. Useful advice to choose properly
sorts of Gaifman graphs, thresholds, and interval widths
remains to be found. After all, parameter tuning is
a black art in many data mining approaches.

One natural variant consists of combining
the constraints defining clans with those of standard 
frequent-set mining; we explored that avenue
and, unfortunately, in all our attempts, we never
found a single case of nontrivial clans.

Also, 
%the
%variants on graphs hinted at in Section~\ref{s:addcons}
%must be developed and their usefulness tested;
%\section{Additional considerations}
%\label{s:addcons}
%Of course, 
we can run this sort of processes on 
multirelational datasets or, even,
directly on graphs. 
For the first case, our examples so far 
fall into the very common and standard 
``single table'' perspective. However, 
from their earliest inception, Gaifman 
graphs were a multirelational concept 
by essence. Applying tree decompositions 
of generalized Gaifman graphs to 
multirelational datasets is, therefore, 
conceptually immediate, and indeed our
example in \ref{fg:uwexponential} comes
from a well-known multirelational benchmark.
However, there, we have not taken into 
account the foreign key phenomenon: would
it be appropriate to denormalize before
computing the Gaifman graph? If so, can
one compute the graph directly, efficiently,
without actually denormalizing the data?

For the second case, graphs are, so to speak, 
their own Gaifman graph, so we can simply 
apply the tree decomposition on the given graph.
A couple of extra possibilities naturally arise.
For instance, we could decompose a 2-structure where
the equivalence classes come from the lengths of the
shortest paths between vertices, or from thresholding
these lengths; or from the vertex- or edge-connectivity
(equivalently, min-cuts, by Menger's theorem), again
possibly thresholded.
% To be developed - ZKCC?
Along this line, there may be interesting 
connections with the topic known as ``blockmodeling''
in social networks, which uses a notion similar to that
of clan, although relaxed through allowing exceptions.

The multiplicity-based generalizations we
have proposed are quite basic; more sophisticate
approaches to define the equivalence relation between
edges might be advantageous. In particular, 
we believe now that some advances might come from 
the study of the applicability of unsupervised 
discretization methods~\cite{DBLP:conf/icml/DoughertyKS95}. 
Indeed, the actual
multiplicities appearing as labels of the edges 
of the Gaifman graph form a set of integers
that is to be discretized in a number of intervals
in an unsupervised manner. A few existing algorithms
for unsupervised discretization can be applied to 
try and automatize parts of the transformation of
the labeled Gaifman graph into the starting 2-structure.

\looseness=-1
Besides the theoretical developments, improving
the software tool is also a desirable endeavor.
Initially, we found the very notion of exploratory 
data analysis via 2-structure decompositions 
of quantitative versions of Gaifman graphs risky
enough, and were not eager to compute very fast, 
nor in a very usable way by other people, results that, 
in principle, were candidates to be fully useless. 
However, we found our initial results 
%(mostly as we have described here)
clearly sufficient to consider that this 
approach is worth of further effort: we did design 
better algorithms than the ones initially
employed \cite{LRodriguez2017}, and we are confident 
that our tool will see considerable improvements 
along several facets in the coming months: the
exploration of alternative tree visualizations, 
the implementation of additional control like 
zooming, or the possibility of importing the data 
directly from databases; this last extension
is actually crucial in order to try our methods on
the usual multirelational benchmarks.

\bibliographystyle{splncs03}
% \bibliography{bibfile}

\begin{thebibliography}{10}
\providecommand{\url}[1]{\texttt{#1}}
\providecommand{\urlprefix}{URL }

%\bibitem{DBLP:conf/ac/BadouelD96}
%Badouel, E., Darondeau, P.: Theory of regions. In: Reisig, W., Rozenberg, G.
%  (eds.) Lectures on Petri Nets {I:} Basic Models, Advances in Petri Nets, the
%  volumes are based on the Advanced Course on Petri Nets, held in Dagstuhl,
%  September 1996. Lecture Notes in Computer Science, vol. 1491, pp. 529--586.
%  Springer (1996), \url{https://doi.org/10.1007/3-540-65306-6_22}

\bibitem{Dawson1995}
Dawson, R.J.M.: The 'unusual episode' data revisited. Journal of Statistics
  Education  3(3) (1995)

\bibitem{UCI}
Dheeru, D., Karra~Taniskidou, E.: {UCI} machine learning repository (2017),
  \url{http://archive.ics.uci.edu/ml}

\bibitem{DBLP:conf/icml/DoughertyKS95}
Dougherty, J., Kohavi, R., Sahami, M.: Supervised and unsupervised
  discretization of continuous features. In: Prieditis, A., Russell, S.J.
  (eds.) Machine Learning, Proceedings of the Twelfth International Conference
  on Machine Learning, Tahoe City, California, USA, July 9-12, 1995. pp.
  194--202. Morgan Kaufmann (1995)

\bibitem{DBLP:books/daglib/0025562}
Ehrenfeucht, A., Harju, T., Rozenberg, G.: The Theory of 2-Structures - {A}
  Framework for Decomposition and Transformation of Graphs. World Scientific
  (1999), \url{http://www.worldscibooks.com/mathematics/4197.html}

\bibitem{DBLP:journals/tcs/EhrenfeuchtR90b}
Ehrenfeucht, A., Rozenberg, G.: Primitivity is hereditary for 2-structures.
  Theor. Comput. Sci.  70(3),  343--358 (1990),
  \url{https://doi.org/10.1016/0304-3975(90)90131-Z}

\bibitem{Gallai1967}
Gallai, T.: Transitiv orientierbare graphen. Acta Mathematica Academiae
  Scientiarum Hungarica  18(1),  25--66 (1967),
  \url{http://dx.doi.org/10.1007/BF02020961}

\bibitem{Graphviz}
Gansner, E.R., North, S.C.: An open graph visualization system and its
  applications to software engineering. SOFTWARE - PRACTICE AND EXPERIENCE
  30(11),  1203--1233 (2000)

\bibitem{DLarraz2010}
Larraz, D.: Aplicaci\'on de las 2-estructuras a las gram\'aticas del lenguaje
  humano y representaci\'on gr\'afica de ambas. Graduation Project, Universidad
  de Zaragoza (2010), \url{http://zaguan.unizar.es/record/5000}

\bibitem{DBLP:books/sp/Libkin04}
Libkin, L.: Elements of Finite Model Theory. Texts in Theoretical Computer
  Science. An {EATCS} Series, Springer (2004),
  \url{https://doi.org/10.1007/978-3-662-07003-1}

\bibitem{DBLP:journals/algorithmica/McConnell95}
McConnell, R.M.: An {O(n{\({^2}\)})} incremental algorithm for modular
  decomposition of graphs and 2-structures. Algorithmica  14(3),  229--248
  (1995), \url{https://doi.org/10.1007/BF01206330}

\bibitem{MCCONNELL1999189}
McConnell, R.M., Spinrad, J.P.: Modular decomposition and transitive
  orientation. Discrete Mathematics  201(1),  189 -- 241 (1999),
  \url{http://www.sciencedirect.com/science/article/pii/S0012365X98003197}

\bibitem{DBLP:journals/corr/abs-1211-3871}
Padhy, N., Panigrahi, R.: Multi relational data mining approaches: {A} data
  mining technique. CoRR  abs/1211.3871 (2012),
  \url{http://arxiv.org/abs/1211.3871}

\bibitem{LRodriguez2017}
Rodr\'\i{}guez-Navas, L.: Estructures de grafs amb equival\`encies d'arestes
  aplicades a l'an\`alisi de dades relacionals. Graduation Project, FIB, UPC
  (2017)

\bibitem{DBLP:books/lib/WittenFH11}
Witten, I.H., Eibe, F., Hall, M.A.: Data mining: practical machine learning
  tools and techniques, 3rd Edition. Morgan Kaufmann, Elsevier (2011),
  \url{http://www.worldcat.org/oclc/262433473}

\end{thebibliography}

\end{document}